\documentclass[pre,twocolumn]{revtex4}
\usepackage{amsmath,amsfonts,epsfig}
\usepackage[english]{babel}
\usepackage{graphicx}
\newcommand{\figwidth}{\columnwidth}

\begin{document}

\title{Poisson-Boltzmann cell model for heterogeneously charged colloids}
\author{Eelco Eggen and Ren\'{e} van Roij}
\affiliation{Institute for Theoretical Physics, Utrecht
University, Leuvenlaan 4, 3584 CE Utrecht, The Netherlands}
\date{7 July 2009}

\begin{abstract}
We introduce the Poisson-Boltzmann cell model for spherical
colloidal particles with a heterogeneous surface charge
distribution. This model is obtained by generalizing existing cell
models for mixtures of homogeneously charged colloidal spheres.
Our new model has similar features as Onsager's second-virial
theory for liquid crystals, but it predicts no orientational
ordering if there is no positional ordering. This implies that all
phases of heterogeneously charged colloids that are liquid-like
with respect to translational degrees of freedom are also
isotropic with respect to particle orientation.
\end{abstract}

\maketitle

\section{Introduction}
Already a long time ago \citet{Marcus}, \citet{Ohtsuki} and
\citet{Alexander} realized that the cell model approach of
\citet{Wigner} to calculate the properties of electrons in solids
can also be applied to colloidal matter. In this case there are no
quantum effects, and in stead of a wave function one calculates
the ion-distributions around charged colloidal particles dispersed
in a liquid medium. In the simplest case, the colloidal dispersion
consists of one colloidal species immersed in a 1:1 electrolyte
solution. The colloidal particles are homogeneously charged and
have a spherical shape. Additionally, the simplification of taking
a spherical Wigner-Seitz cell|instead of space filling|is
justified in fluid phases with no broken translational symmetry.

A number of extensions has been made to this basic cell model.
There is the eccentric Poisson-Boltzmann cell model
\citep{Ohtsuki,Gruenberg:ecc}, and the heterogeneous|or
polydisperse|cell model \citep{Biesheuvel,Torres}, to describe
mixtures. Additionally, the cell model has been extended by
applying cylindrical Wigner-Seitz cells for the description of
disc-shaped and rodlike particles \citep{Hansen,Bocquet}. In the
present paper, we make an extension towards particles with an
heterogeneous surface charge distribution, such as patterned
colloids \citep{Glotzer} or Janus particles \citep{Walther}. Janus
particles are characterized by two distinct regions of surface
area. Each of these ``faces'' has a different chemical
composition, which can create spontaneous aggregation
\citep{Hong}. Our aim is towards a description of
self-organization of these particles from single-particle
properties. The cell model can be a powerful tool, giving a simple
description of these complex systems. From this description, a
number of thermodynamic quantities can be derived such as free
energy, and osmotic pressure.

We investigate how surface charge heterogeneity influences the
distribution of particle orientations in the case of a homogeneous
positional distribution. This case can be considered as a
simplified description of a fluid of these particles, but also as
an over-simplified description of a solid. Since the particle
interactions are implemented through the boundary conditions of
the Wigner-Seitz cell, one can choose to neglect the correlations
between the orientational and positional ordering. The basis of
our model is the generalization of the Poisson-Boltzmann cell
model by \citet{Biesheuvel} considering mixtures, together with
the insight of \citet{Onsager} that a distribution of orientations
can be considered in the same way as mixtures. In principle, this
model can be used to predict the phase behavior of a large class
of colloidal and nanoparticles, because an anisotropic (as well as
spherical) particle shape can be treated. However, here we
restrict ourselves to a system of spherical colloidal particles.

\begin{figure}[ht]
\includegraphics[width=0.8\figwidth]{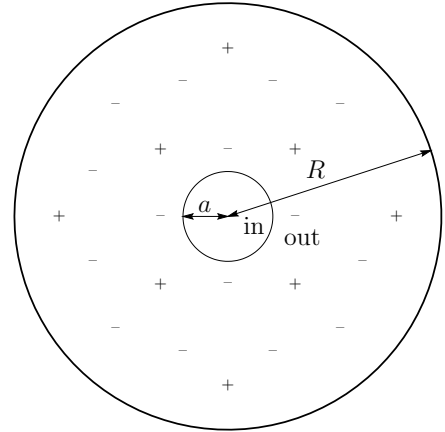}
\caption{Illustration of the Poisson-Boltzmann cell model. A
colloidal particle of radius $a$ is surrounded by ions. It is
situated in the center of a spherical cell of radius
$R$.}\label{fig:cell-dim}
\end{figure}

\section{The cell model for mixtures}
We start by giving the prerequisites for the ordinary spherically
symmetric cell model (see Fig.~\ref{fig:cell-dim}). In this simple
case, we consider a system of $N$ identical, spherically symmetric
colloidal particles in a volume $V$. These particles have radius
$a$ and a surface charge $Ze$, where $e$ is the elementary charge.
The system is presumed to be in osmotic contact with a reservoir
of monovalent cations and anions at total concentration
$2\rho_{\rm s}$, with charge $\pm e$. Both the particle and the
surrounding medium are considered to be dielectric media, where
$\epsilon_{\rm in}$ and $\epsilon_{\rm out}$ are the dielectric
constants of the particle and the solvent, respectively.

Within Poisson-Boltzmann theory, one relates the denisty profiles
$\rho_{\pm}(\mathbf{r})$ of the ions to the electrostatic
potential $\Psi(\mathbf{r})$ in a fixed configuration of the
colloidal particles. This complicated nonlinear $N$-body problem
can be simplified considerably by considering a single colloidal
particle in the center of a Wigner-Seitz cell, surrounded by the
ions. This cell is assumed to have a spherical shape (of radius
$R$) instead of space filling. The volume of the cell is fixed by
the available volume per particle,
\begin{equation}
\frac{4\pi}{3}R^{3} = \frac{V}{N}.
\end{equation}
We treat the ions in a mean field description, such that we obtain
the Poisson-Boltzmann (PB) equation
\begin{equation}\label{eq:PB}
\nabla^{2}\Phi(\mathbf{r}) = \left\{
\begin{array}{ll}
0 & \mbox{ for $0<|\mathbf{r}|<a$,}\\
\kappa^{2}\sinh\Phi(\mathbf{r}) & \mbox{ for $a<|\mathbf{r}|<R$,}
\end{array}\right.
\end{equation}
where $\kappa^{-1}=\sqrt{8\pi\epsilon_{\rm out}/\rho_{\rm s}\beta
e^{2}}$ is the Debye screening length, and $\Phi=\beta e\Psi$ is
the dimensionless electrostatic potential, where $\beta=1/k_{\rm
B}T$. This second-order nonlinear partial differential equation
describes the electrostatic effects of the ions in the cell volume
surrounding the colloidal particle. The boundary conditions are
determined by the charge on the particle and the cell
electroneutrality. The first boundary condition is given by
\begin{equation}
\Phi(\mathbf{r})\Bigr|_{r\uparrow a} =
\Phi(\mathbf{r})\Bigr|_{r\downarrow a},
\end{equation}
\begin{equation}\label{eq:boun:col-homo}
\epsilon_{\rm in}\frac{\partial}{\partial
r}\Phi(\mathbf{r})\biggr|_{r\uparrow a} = \beta e^{2}\frac{Z}{4\pi
a^{2}} + \epsilon_{\rm out}\frac{\partial}{\partial
r}\Phi(\mathbf{r})\biggr|_{r\downarrow a},
\end{equation}
which fixes the electric field at the particle surface. In
Eq.~\eqref{eq:boun:col-homo}, $Z$ is the number of elementary
charges on the particle surface. Because of the homogeneous
surface charge distribution, there is no electric field inside the
particle. This reduces boundary condition \eqref{eq:boun:col-homo}
to
\begin{equation}\label{eq:boun:col-E-homo}
\frac{\partial}{\partial r}\Phi(\mathbf{r})\biggr|_{r\downarrow a}
\equiv \Phi^{\prime}(a) = -\frac{Zl_{\rm B}}{a^{2}},
\end{equation}
where $l_{\rm B}=\beta e^{2}/4\pi\epsilon_{\rm out}$ is the
Bjerrum length. The second boundary condition fixes the electric
field at the cell boundary, according to Gauss' law
\begin{equation}\label{eq:boun:cell-E-homo}
\Phi^{\prime}(R) = 0.
\end{equation}
The PB-equation \eqref{eq:PB} together with the boundary
conditions \eqref{eq:boun:col-E-homo} and
\eqref{eq:boun:cell-E-homo} form a closed set and describe the
basic Poisson-Boltzmann cell model as studied for example in
Refs.~\citep{Alexander,Gruenberg:gas,Trizac:cell}.

Now we discuss the case of a mixture of equally sized,
homogeneously charged colloidal species with surface charge
$Z_{i}e$. Again, the particles have radius $a$, and are positioned
in the center of a spherical Wigner-Seitz cell of radius $R$. The
PB-equation \eqref{eq:PB} is solved separately for each colloidal
species $i$, and the notion of electroneutrality is applied to the
system as a whole \citep{Biesheuvel}. Each solution $\Phi_{i}(r)$
is determined by the surface charge density on the corresponding
colloidal species
\begin{equation}\label{eq:boun:col-E-iso}
\Phi_{i}^{\prime}(a) = -\frac{Z_{i}l_{\rm B}}{a^{2}}.
\end{equation}
Given that the cells of each pair of species are considered to be
neighboring, it is natural to impose a common boundary value
$\Phi_{R}$ for the potential on every cell surface. This value is
by definition equal to the average value of the electrostatic
potential at the cell boundary for different species. Hence, we
have
\begin{equation}\label{eq:boun:cell-V-iso}
\Phi_{i}(R) = \sum_{j}x_{j}\Phi_{j}(R) \equiv \Phi_{R}
\qquad\forall i,
\end{equation}
where $x_{i}=N_{i}/N$ is the molar fraction of species $i$, such
that
\begin{equation}
\sum_{i}x_{i} = 1.
\end{equation}
The boundary value $\Phi_{R}$ is fixed by setting the average
value of the charge contained in the cells for different species
to zero. According to Gauss' law, this is achieved by imposing
\begin{equation}\label{eq:boun:cell-E-iso}
\sum_{i}x_{i}\Phi_{i}^{\prime}(R) = 0.
\end{equation}
The PB-equation \eqref{eq:PB} applied to each species $i$ together
with the boundary conditions \eqref{eq:boun:col-E-iso},
\eqref{eq:boun:cell-V-iso}, and \eqref{eq:boun:cell-E-iso} again
form a closed set.

In principle, one could allow for a different cell radius $R_{i}$
for each colloidal species. In this case, a factor $R_{i}^{2}$
must be included in the summation of
Eq.~\eqref{eq:boun:cell-E-iso}. One subsequently imposes a set of
physically motivated conditions on these radii. These conditions
must comply with the fact that the average cell volume of the
system is given by
\begin{equation}
\frac{4\pi}{3}\sum_{i}x_{i}R_{i}^{3} = \frac{V}{N}.
\end{equation}
In the case of mixtures where the charge of all species has the
same sign, one can fix each $R_{i}$ by imposing the condition that
the electric field must vanish at the cell boundary of each
species \citep{Torres}. For reasons of simplicity, however, we
will not use such an extension. Hence, in this paper, the cell
radius is given by a single value $R$.

\section{An extension towards heterogeneous charge distributions}
We now consider a system of spherical colloidal particles that are
\emph{not} homogeneously charged, such as Janus particles
\citep{Walther}. Again, the system consists of $N$ identical
spherical particles of radius $a$ in a volume $V$. We apply the
same Poisson-Boltzmann theory, such that the electrostatic
potential obeys the PB-equation \eqref{eq:PB}. As before, we fix
the position of each particle at the center of a spherical
Wigner-Seitz cell of radius $R$, and the PB-equation must be
solved for each ``species''. However, in this case we replace the
species index $i$ by an orientation $\hat{\omega}$, and each
solution is denoted by $\Phi(\hat{\omega};\mathbf{r})$. In the
spirit of Onsager, we can view such an orientational distribution
as a mixture where the different particle species have a distinct
charge distribution.

In this paper, we focus on charge distributions that are
cylindrically symmetric with respect to the particle orientation
$\hat{\omega}$. This charge distribution gives us the boundary
condition on the particle surface at each position
$a\hat{\mathbf{n}}$
\begin{equation}\label{eq:boun:col-V}
\Phi_{\rm in}(\hat{\omega};a\hat{\mathbf{n}}) = \Phi_{\rm
out}(\hat{\omega};a\hat{\mathbf{n}}),
\end{equation}
\begin{equation}\label{eq:boun:col-E}
\epsilon\,\hat{\mathbf{n}}\cdot\nabla\Phi_{\rm
in}(\hat{\omega};a\hat{\mathbf{n}}) = 4\pi l_{\rm
B}\sigma(\hat{\omega};\hat{\mathbf{n}}) +
\hat{\mathbf{n}}\cdot\nabla\Phi_{\rm
out}(\hat{\omega};a\hat{\mathbf{n}}),
\end{equation}
where $\sigma(\hat{\omega};\hat{\mathbf{n}})$ is the surface
charge density (in units of $e$) that belongs to a particle with
an orientation $\hat{\omega}$, and $\epsilon=\epsilon_{\rm
in}/\epsilon_{\rm out}$ is the relative dielectric constant of the
particle with respect to the surrounding solvent. The labels
``in'' and ``out'' denote the solutions inside the colloidal
particle, and outside the particle, respectively.

\begin{figure}[t]
\includegraphics[width=\figwidth]{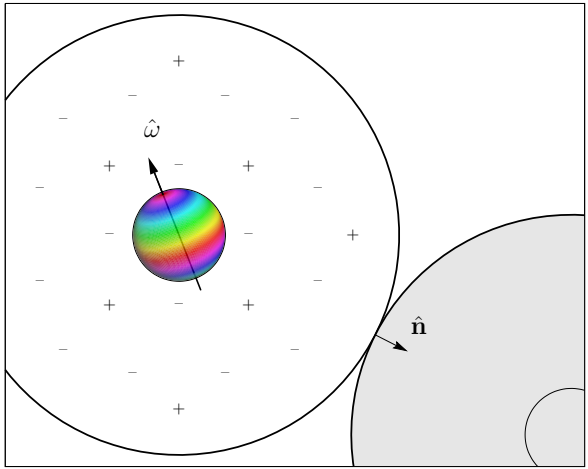}
\caption{Illustration of the Poisson-Boltzmann cell model for
heterogeneous charge distributions. For each direction
$\hat{\mathbf{n}}$ perpendicular to the cell surface, we determine
the appropriate boundary conditions.}\label{fig:cell-boun}
\end{figure}

Next, we must supply a generalized version of the boundary
conditions on the cell surface given in
Eqs.~\eqref{eq:boun:cell-V-iso} and \eqref{eq:boun:cell-E-iso}.
The following approach is illustrated by Fig.~\ref{fig:cell-boun},
which shows the directionality that must be included in the
appropriate boundary conditions. First, the concept of a fixed
cell surface potential is applied to include different positions
on the cell surface. Similar to the cell model for mixtures, the
value of this potential is defined as the average value of the
electrostatic potential at the cell boundary for different
orientations. However, the average value is taken at the position
of opposite orientation
\begin{equation}\label{eq:boun:cell-V-ani}
\Phi(\hat{\omega};R\hat{\mathbf{n}}) =
\langle\Phi\rangle(-R\hat{\mathbf{n}}) \equiv
\Phi_{R}(-\hat{\mathbf{n}}) \qquad\forall\hat{\omega},
\end{equation}
where instead of a summation over species weighted by the molar
fractions $x_{i}$, we have an integral over particle orientations
$\hat{\omega}$ weighted by the orientational distribution function
(ODF) $f(\hat{\omega})$
\begin{equation}\label{eq:ODF:ave}
\langle\Phi\rangle(\mathbf{r}) \equiv \int{\rm
d}\hat{\omega}f(\hat{\omega})\Phi(\hat{\omega};\mathbf{r}).
\end{equation}
This distribution is normalized such that
\begin{equation}\label{eq:ODF:norm}
\int{\rm d}\hat{\omega}f(\hat{\omega}) = 1.
\end{equation}
The definition of the cell surface potential
$\Phi_{R}(\hat{\mathbf{n}})$ in Eq.~\eqref{eq:boun:cell-V-ani} is
such that it is always an even function
\begin{equation}
\Phi_{R}(\hat{\mathbf{n}}) = \Phi_{R}(-\hat{\mathbf{n}}).
\end{equation}

Finally, we have to impose a boundary condition that fixes this
cell surface potential. However, if we only impose global
electroneutrality on the system, we obtain a boundary condition
that is too general for a solution that is \emph{not} spherically
symmetric. It would ensure that the average value (over all
``species'') of the charge contained in each Wigner-Seitz cell
vanishes. By applying Gauss' law, we see that this condition is
satisfied by setting the average value of the electric field
integrated over the surface to zero
\begin{equation}
\int{\rm
d}\hat{\mathbf{n}}\bigl[\hat{\mathbf{n}}\cdot\nabla\langle\Phi\rangle(R\hat{\mathbf{n}})\bigr]
= 0.
\end{equation}
Interestingly, this fixes only the isotropic contribution to the
cell surface potential. Therefore, we impose an additional
condition that is based on the concept of continuity of the
electric field flux from one cell to another. The difference
between the outward flux at the cell boundary and the average
inward flux of neighboring cells is represented by
\begin{equation}
\Delta F_{R}(\hat{\omega};\hat{\mathbf{n}}) \equiv
\hat{\mathbf{n}}\cdot\nabla\Phi(\hat{\omega};R\hat{\mathbf{n}}) -
\hat{\mathbf{n}}\cdot\nabla\langle\Phi\rangle(-R\hat{\mathbf{n}}).
\end{equation}
This quantity is averaged over all particle orientations, and set
to zero, in order to insure global electric field flux
conservation
\begin{equation}
\langle\Delta F_{R}\rangle(\hat{\mathbf{n}}) = 0,
\end{equation}
which is equivalent to imposing
\begin{equation}\label{eq:boun:cell-E-ani}
\hat{\mathbf{n}}\cdot\nabla\langle\Phi\rangle(R\hat{\mathbf{n}}) =
\hat{\mathbf{n}}\cdot\nabla\langle\Phi\rangle(-R\hat{\mathbf{n}}).
\end{equation}
This condition does fix the cell surface potential, and it defines
an average boundary value of the radial derivative such that it is
an odd function of $\hat{\mathbf{n}}$.

\section{Special limiting cases: perfectly isotropic and perfectly aligned}
In this section, we apply specific choices for the ODF. In turn,
these choices yield a specific form for the boundary conditions
\eqref{eq:boun:cell-V-ani} and \eqref{eq:boun:cell-E-ani}. The
resulting models are less intricate than the full model we
presented in the previous section. Also, these models yield
boundary conditions that one would expect from a naive description
of such systems.

Let us take a look at the model that our boundary conditions
yields when we implement specific ODFs. First, we consider a
perfectly isotropic orientational distribution
\begin{equation}\label{eq:f-iso}
f_{\rm iso}(\hat{\omega}) = \frac{1}{4\pi}.
\end{equation}
Since in such a system there is no preferential direction, we
argue that all solutions|for different particle orientations|are
equivalent. Consequently, the cell surface potential
$\Phi_{R}(\hat{\mathbf{n}})$ is independent of the position on the
cell surface
\begin{equation}
\Phi_{\rm iso}(\hat{\omega};R\hat{\mathbf{n}}) = \Phi_{R}.
\end{equation}
This result is in accordance with the notion that in the isotropic
case the average over all particle orientations \eqref{eq:ODF:ave}
is equal to the average over all orientations $\hat{\mathbf{n}}$
of the position on the cell surface, and that this average no
longer depends on either orientation. Therefore, the boundary
condition \eqref{eq:boun:cell-E-ani} is equivalent to the
condition that each cell is electroneutral
\begin{equation}
\hat{\mathbf{n}}\cdot\nabla\langle\Phi_{\rm
iso}\rangle(R\hat{\mathbf{n}}) = \int{\rm
d}\hat{\mathbf{n}}\bigl[\hat{\mathbf{n}}\cdot\nabla\Phi_{\rm
iso}(\hat{\omega};R\hat{\mathbf{n}})\bigr] = 0.
\end{equation}

Alternatively, we can choose a perfectly aligned orientational
distribution
\begin{equation}
f_{\parallel}(\hat{\omega}) =
\delta(\hat{\omega}-\hat{\mathbf{z}}).
\end{equation}
Clearly, in this case there is only one solution to be determined
\begin{equation}
\Phi_{\parallel}(\hat{\mathbf{z}};\mathbf{r}) \equiv
\Phi_{\parallel}(\mathbf{r}),
\end{equation}
and the boundary conditions \eqref{eq:boun:cell-V-ani} and
\eqref{eq:boun:cell-E-ani} read
\begin{equation}
\Phi_{\parallel}(R\hat{\mathbf{n}}) =
\Phi_{\parallel}(-R\hat{\mathbf{n}}),
\end{equation}
\begin{equation}
\hat{\mathbf{n}}\cdot\nabla\Phi_{\parallel}(R\hat{\mathbf{n}}) =
\hat{\mathbf{n}}\cdot\nabla\Phi_{\parallel}(-R\hat{\mathbf{n}}).
\end{equation}
Evidently, this choice leads to periodic boundary conditions.

\section{Application to linearized Poisson-Boltzmann theory}
To solve the full nonlinear problem is possible numerically, but
it turns out to be very involved \citep{Boon}. Therefore, we
restrict ourselves to the linearized version of Poisson-Boltzmann
theory. In this case, the nonlinear right hand side of the
PB-equation is linearized around a certain value. We denote it by
$\Phi_{0}$, such that the linearized Poisson-Boltzmann (LPB)
equation is given by
\begin{equation}\label{eq:LPB}
\nabla^{2}\Phi_{\rm out}(\hat{\omega};\mathbf{r}) =
\kappa^{2}\cosh\Phi_{0}(\Phi_{\rm
out}(\hat{\omega};\mathbf{r})-\Phi_{0}) + \kappa^{2}\sinh\Phi_{0}.
\end{equation}
In some cases, the value for $\Phi_{0}$ is chosen to be zero. This
choice is meaningful if the concentration of colloids, as well as
the total surface charge density, is low. Alternatively, its value
can be set to the isotropically averaged value of the potential at
the cell boundary. This choice is particularly useful when one has
the boundary values of the potential and the electric field from
numerical calculations of the nonlinear PB-equation \citep{Boon}.
These can be used to fit renormalized charge distributions on the
particle surface using the expression in Eq.~\eqref{eq:sol-V}.
Lastly, one can apply the Donnan potential as the value around
which to perform the linearization. This value requires no other
input than the colloid concentration, its particle radius, its
total surface charge, and the reservoir salt concentration
\citep{Gruenberg:gas}. In this paper, we leave $\Phi_{0}$
unspecified.

Inside the colloidal particle $\Phi$ still satisfies the Laplace
equation. It is natural in this case to expand both the inner and
the outer solution in spherical harmonics. This leads to two sets
of coefficients which have to be matched at the particle surface.
Inside the particle
\begin{equation}\label{eq:sol-in}
\Phi_{\rm in}(\hat{\omega};\mathbf{r}) =
\sum_{\ell=0}^{\infty}\sum_{m=-\ell}^{+\ell}A_{\ell,m}(\hat{\omega})r^{\ell}Y_{\ell,m}(\theta,\phi),
\end{equation}
whereas in the cell interior
\begin{align}\label{eq:sol-out}
&\Phi_{\rm out}(\hat{\omega};\mathbf{r}) = \Phi_{0} - \tanh\Phi_{0}\nonumber\\
&{}+
\sum_{\ell=0}^{\infty}\sum_{m=-\ell}^{+\ell}[B_{\ell,m}(\hat{\omega})i_{\ell}(\bar{\kappa}r)
+
C_{\ell,m}(\hat{\omega})k_{\ell}(\bar{\kappa}r)]Y_{\ell,m}(\theta,\phi),
\end{align}
where $\bar{\kappa}^{2}=\kappa^{2}\cosh\Phi_{0}$, and $i_{\ell}$
and $k_{\ell}$ are the modified spherical Bessel functions of
order $\ell$ of the first and second kind, respectively. The
boundary condition on the particle surface are given by
Eqs.~\eqref{eq:boun:col-V} and \eqref{eq:boun:col-E}, where we
decompose the charge distribution as
\begin{equation}\label{eq:charge-dist}
\sigma(\hat{\omega};\hat{\mathbf{n}}) =
\sum_{\ell=0}^{\infty}\frac{2\ell+1}{4\pi}\sigma_{\ell}P_{\ell}(\hat{\omega}\cdot\hat{\mathbf{n}}).
\end{equation}
Next, we impose the boundary conditions at the cell surface, which
are given in Eqs.~\eqref{eq:boun:cell-V-ani} and
\eqref{eq:boun:cell-E-ani}. Together, this yields the general
solution for the dimensionless electrostatic potential in the cell
interior
\begin{align}\label{eq:sol-V}
&\Phi_{\rm out}(\hat{\omega};\mathbf{r}) = \Phi_{0} -
\tanh\Phi_{0} + l_{\rm B}\bar{\kappa}^{-1}
\sum_{\ell=0}^{\infty}\frac{(2\ell+1)\sigma_{\ell}}{\Xi_{\ell}(\epsilon;\bar{\kappa}a,\bar{\kappa}R)}\nonumber\\
&{}\times
\bigl[k_{\ell}(\bar{\kappa}r)i_{\ell}(\bar{\kappa}R)-i_{\ell}(\bar{\kappa}r)k_{\ell}(\bar{\kappa}R)\bigr]
P_{\ell}(\hat{\omega}\cdot\hat{\mathbf{r}})\nonumber\\
&{}+ l_{\rm B}\bar{\kappa}^{-1}
\underbrace{\sum_{\ell=0}^{\infty}}_{\ell\;{\rm
even}}\frac{(2\ell+1)\sigma_{\ell}}
{\Lambda_{\ell}(\epsilon;\bar{\kappa}a,\bar{\kappa}R)\bar{\kappa}^{2}R^{2}\Xi_{\ell}(\epsilon;\bar{\kappa}a,\bar{\kappa}R)}\nonumber\\
&{}\times \Xi_{\ell}(\epsilon;\bar{\kappa}a,\bar{\kappa}r)\int{\rm
d}\hat{\omega}^{\prime}f(\hat{\omega}^{\prime})P_{\ell}(\hat{\omega}^{\prime}\cdot\hat{\mathbf{r}}),
\end{align}
where
\begin{align}\label{eq:Xi}
\Xi_{\ell}(\epsilon;\bar{\kappa}a,\bar{\kappa}R) \equiv{}&
-\left(k_{\ell}^{\prime}(\bar{\kappa}a) -
\frac{\epsilon\ell}{\bar{\kappa}a}k_{\ell}(\bar{\kappa}a)\right)i_{\ell}(\bar{\kappa}R)\nonumber\\
&{}+ \left(i_{\ell}^{\prime}(\bar{\kappa}a) -
\frac{\epsilon\ell}{\bar{\kappa}a}i_{\ell}(\bar{\kappa}a)\right)k_{\ell}(\bar{\kappa}R),\\
\label{eq:Lambda}
\Lambda_{\ell}(\epsilon;\bar{\kappa}a,\bar{\kappa}R) \equiv{}&
\frac{\partial\Xi_{\ell}(\epsilon;\bar{\kappa}a,\bar{\kappa}R)}{\partial(\bar{\kappa}R)}.
\end{align}
The details of the derivation of the expression in
Eq.~\eqref{eq:sol-V} can be found in appendix \ref{sec:deriv}.
Note that the first sum (over all $\ell$) does not depend on the
ODF, whereas it does depend on the particle orientation
$\hat{\omega}$. This contribution to the potential is purely due
to the particle at the center of the cell, and it vanishes at the
cell boundary. Conversely, the second sum (over even $\ell$) does
not depend on the particle orientation, whereas it does depend on
the ODF. This means that it describes the effect of all the
surrounding particles. Moreover, it vanishes in the limit of
infinite dilution ($R\rightarrow\infty$).

The thermodynamic potential for the ion distribution in a single
cell is given by \citep{Gruenberg:gas}
\begin{align}
\beta\Omega_{\rm cell}(\hat{\omega}) ={}& \rho_{\rm s}\int_{\rm
out}{\rm
d}\mathbf{r}\bigl\{\Phi_{\rm out}(\hat{\omega};\mathbf{r})\sinh[\Phi_{\rm out}(\hat{\omega};\mathbf{r})]\nonumber\\
&\qquad\qquad{}- 2\cosh[\Phi_{\rm out}(\hat{\omega};\mathbf{r})] +
2\bigr\}\nonumber\\
&{}+ \frac{a^{2}}{2}\int{\rm
d}\hat{\mathbf{n}}\sigma(\hat{\omega};\hat{\mathbf{n}})\Phi_{\rm
out}(\hat{\omega};a\hat{\mathbf{n}}),
\end{align}
where the the label ``out'' at the integral symbol denotes
integration over the cell interior (i.e., the domain of $\Phi_{\rm
out}$), and $\rho_{\rm s}$ is the reservoir salt concentration
(such that $\kappa^{2}=8\pi l_{\rm B}\rho_{\rm s}$). We cannot
evaluate this expression analytically. Therefore, we linearize it
around $\Phi_{0}$ to find
\begin{equation}\label{eq:Omega}
\beta\Omega_{\rm cell}(\hat{\omega}) \simeq \beta\Omega_{0} +
\beta\Omega_{\rm iso} + \beta\Omega_{\rm int},
\end{equation}
where
\begin{align}
\beta\Omega_{0} \equiv{}& \frac{4\pi}{3}(R^{3}-a^{3})\rho_{\rm s}
(\Phi_{0}\sinh\Phi_{0} - 2\cosh\Phi_{0} + 2),\\
\beta\Omega_{\rm iso} \equiv{}& \rho_{\rm s}(\Phi_{0}\cosh\Phi_{0}
- \sinh\Phi_{0})\nonumber\\
&\qquad\qquad{}\times \int_{\rm out}{\rm d}\mathbf{r}(\Phi_{\rm
out}(\hat{\omega};\mathbf{r})-\Phi_{0}),\\
\beta\Omega_{\rm int} \equiv{}& \frac{a^{2}}{2}\int{\rm
d}\hat{\mathbf{n}}\sigma(\hat{\omega};\hat{\mathbf{n}})\Phi_{\rm
out}(\hat{\omega};a\hat{\mathbf{n}}).
\end{align}
It turns out that $\beta\Omega_{\rm int}$ depends on the particle
orientation and the ODF, whereas the other two terms depend on
neither. The expression we obtain can be derived through another
route, by expanding the original non-linear functional of the ion
profiles $\rho_{\pm}(\hat{\omega};\mathbf{r})$ up to second order
with respect to a density $\rho_{\pm,0}=\rho_{\rm
s}\exp[\mp\Phi_{0}]$. Minimizing this functional with respect to
the ion profiles yields the LPB-equation \eqref{eq:LPB}, and the
accompanying expressions for the ion profiles
$\rho_{\pm}(\hat{\omega};\mathbf{r})\simeq\rho_{\pm,0}[1\pm\Phi_{0}\mp\Phi_{\rm
out}(\hat{\omega};\mathbf{r})]$. Substitution of this expression
into the functional yields Eq.~\eqref{eq:Omega}.

\section{Onsager-like second-order density functional theory}
We approximate the total free energy (per colloidal particle) of
the system by averaging over all particle orientations. Also, we
add an entropic contribution, which is analogous to mixing entropy
\begin{equation}
\frac{\beta\mathcal{F}[f]}{N} \simeq \int{\rm
d}\hat{\omega}f(\hat{\omega})\ln[4\pi f(\hat{\omega})] + \int{\rm
d}\hat{\omega}f(\hat{\omega})\beta\Omega_{\rm cell}(\hat{\omega}).
\end{equation}
We neglect the translational entropic contributions of the
colloidal particles, because we are only interested in the effects
of charge anisotropy and orientational distribution. Using the
identity
\begin{equation}
\int{\rm
d}\hat{\mathbf{n}}\sigma(\hat{\omega};\hat{\mathbf{n}})P_{\ell}(\hat{\omega}^{\prime}\cdot\hat{\mathbf{n}})
= \sigma_{\ell}P_{\ell}(\hat{\omega}\cdot\hat{\omega}^{\prime}),
\end{equation}
which can be easily derived from Eq.~\eqref{eq:charge-dist} using
the addition theorem, we derive the following expression for the
free energy
\begin{align}\label{eq:F}
\frac{\beta\mathcal{F}[f]}{N} \simeq{}&
\frac{\beta\mathcal{F}_{0}}{N} + \int{\rm
d}\hat{\omega}f(\hat{\omega})\ln[4\pi f(\hat{\omega})]\nonumber\\
&{}+ \frac{1}{2}\int{\rm d}\hat{\omega}f(\hat{\omega})\int{\rm
d}\hat{\omega}^{\prime}f(\hat{\omega}^{\prime})K(\hat{\omega},\hat{\omega}^{\prime}),
\end{align}
where
\begin{align}
&\frac{\beta\mathcal{F}_{0}}{N} \equiv \beta\Omega_{0}\nonumber\\
&{}+ (\Phi_{0} - \tanh\Phi_{0})\left(\frac{a^{2}\sigma_{0}}{2} -
\frac{4\pi}{3}(R^{3}-a^{3})\rho_{\rm
s}\sinh\Phi_{0}\right)\nonumber\\
&{}+ \frac{a^{2}\sigma_{0}}{2}(\Phi_{0} - \tanh\Phi_{0}) +
\frac{a^{2}l_{\rm B}\bar{\kappa}^{-1}}{2}
\sum_{\ell=0}^{\infty}\frac{(2\ell+1)\sigma_{\ell}^{2}}{\Xi_{\ell}(\epsilon;\bar{\kappa}a,\bar{\kappa}R)}\nonumber\\
&{}\times
\bigl[k_{\ell}(\bar{\kappa}a)i_{\ell}(\bar{\kappa}R)-i_{\ell}(\bar{\kappa}a)k_{\ell}(\bar{\kappa}R)\bigr],
\end{align}
and
\begin{equation}\label{eq:K}
K(\hat{\omega},\hat{\omega}^{\prime}) \equiv \frac{l_{\rm
B}}{\bar{\kappa}^{5}R^{2}}\underbrace{\sum_{\ell=0}^{\infty}}_{\ell\;{\rm
even}}\frac{(2\ell+1)\sigma_{\ell}^{2}P_{\ell}(\hat{\omega}\cdot\hat{\omega}^{\prime})}
{\Lambda_{\ell}(\epsilon;\bar{\kappa}a,\bar{\kappa}R)\Xi_{\ell}(\epsilon;\bar{\kappa}a,\bar{\kappa}R)}.
\end{equation}
Note that the second term of $\beta\mathcal{F}_{0}/N$ vanishes if
one chooses the Donnan potential for the value of $\Phi_{0}$.
Also, the entire contribution from $\beta\mathcal{F}_{0}/N$
vanishes upon taking the functional derivative with respect to
$f(\hat{\omega})$.

The structure of the free energy functional \eqref{eq:F} is
remarkably similar to that of Onsager's second-virial theory for
hard rods \citep{Onsager}. In the case of spherocylinders, the
kernel $K(\hat{\omega},\hat{\omega}^{\prime})$ stems from
hard-core interactions, and it is equal to the product of the rod
density and the orientation dependent excluded volume of one rod
in the vicinity of another,
\begin{equation}
K(\hat{\omega},\hat{\omega}^{\prime}) =
\frac{N}{V}\left[\frac{4\pi}{3}D^{3} + 2\pi L^{2}D +
2L^{2}D\sin\gamma\right],
\end{equation}
where $L$ is the rod length, $D$ is the rod diameter. The angle
$\gamma\in[0,\pi]$ between the two rod orientations is defined by
$\cos\gamma\equiv\hat{\omega}\cdot\hat{\omega}^{\prime}$.
Onsager's second-virial theory predicts the existence of an
isotropic--nematic transition, caused by the competition between
orientational and translational entropy. In the low-density
isotropic phase, the ``mixing'' term $\int{\rm
d}\hat{\omega}f(\hat{\omega})\ln[4\pi f(\hat{\omega})]$ is
minimized by an isotropic orientational distribution, whereas the
contribution due to the average excluded volume
$\frac{1}{2}\int{\rm d}\hat{\omega}f(\hat{\omega})\int{\rm
d}\hat{\omega}^{\prime}f(\hat{\omega}^{\prime})K(\hat{\omega},\hat{\omega}^{\prime})$
is minimized in the high-density nematic phase. This transition
only occurs if the length-to-diameter $L/D$ is large, such that
the kernel $K(\hat{\omega},\hat{\omega}^{\prime})$ is sufficiently
anisotropic, with a maximum at $\gamma=\pi/2$. In the limit $L\gg
D$, the description by Onsager is quantitative.

\begin{figure}[t]
\includegraphics[width=\figwidth]{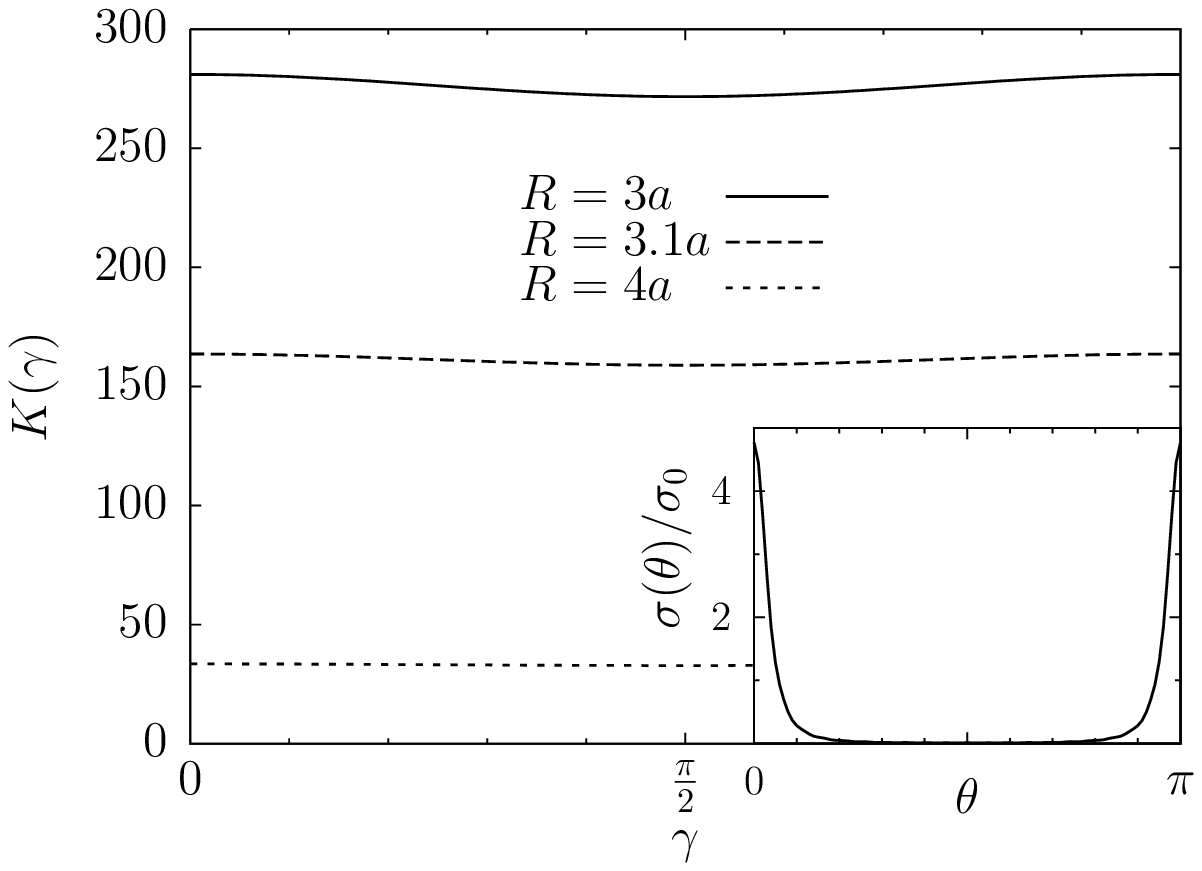}
\includegraphics[width=\figwidth]{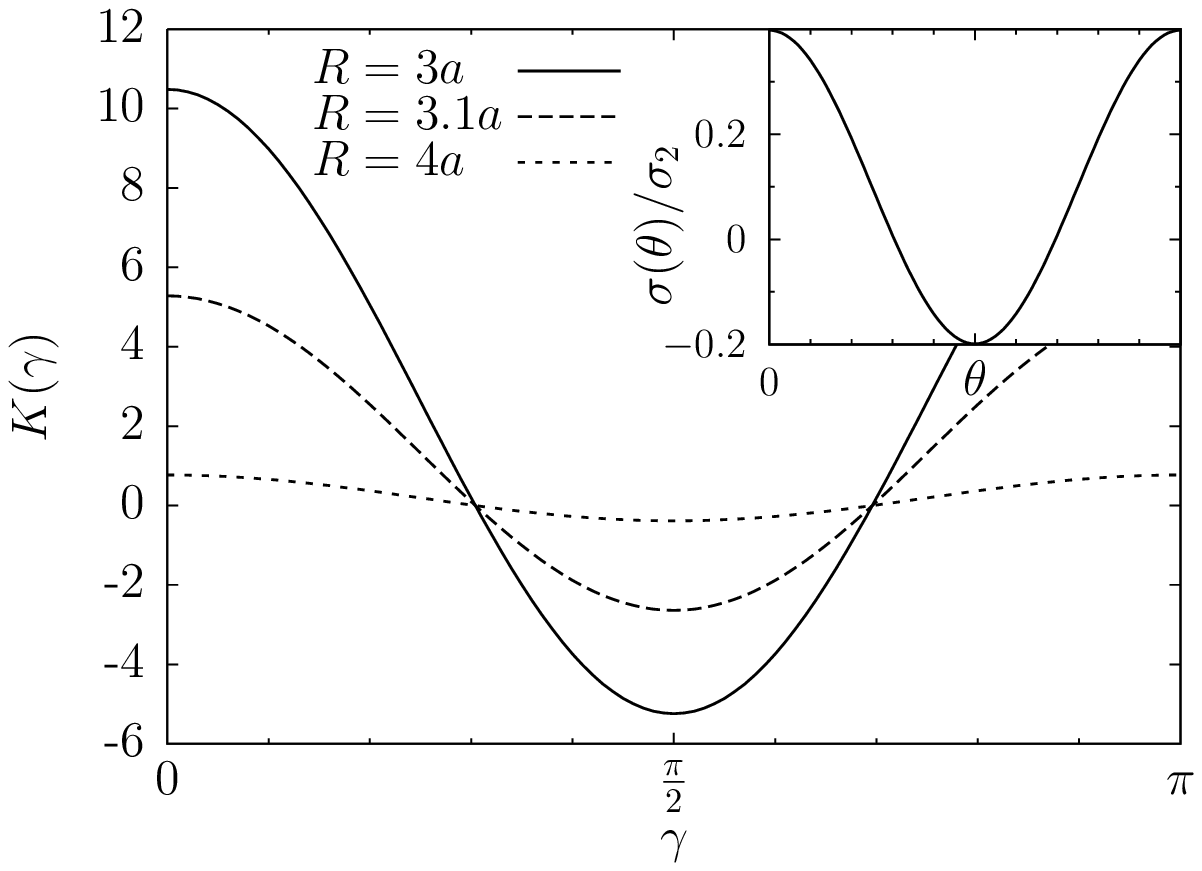}
\caption{The kernel function
$K(\hat{\omega},\hat{\omega}^{\prime})$ for different values of
the cell radius $R$, and two distinct charge distributions. We
fixed the values $\bar{\kappa}a=1$, $\epsilon=1$, and
$\bar{\kappa}l_{\rm B}=0.01$. The inset in each graph shows the
corresponding charge distribution as a function of the angle
$\theta$ between the axis of symmetry and the position vector on
the surface. The charge distribution in the top graph is scaled
with $\sigma_{0}=10^{3}\bar{\kappa}^{2}$, whereas the scaling in
the lower graph is given by $\sigma_{2}=10^{3}\bar{\kappa}^{2}$.}
\label{fig:kernel}
\end{figure}

In the present case, the kernel stems from anisotropic
electrostatic interactions, and we will investigate if these can
give rise to such a symmetry-breaking transition. In both cases,
the kernel is rotationally invariant, i.e., it only depends on the
mutual relative orientation of the unit vectors $\hat{\omega}$ and
$\hat{\omega}^{\prime}$|through the dot product
$\hat{\omega}\cdot\hat{\omega}^{\prime}$. Figure \ref{fig:kernel}
shows the values of the kernel
$K(\hat{\omega},\hat{\omega}^{\prime})$ for two distinct surface
charge distributions $\sigma(\hat{\omega},\hat{\mathbf{n}})$. The
angle $\theta$ between the axis of symmetry and the position
vector on the surface is defined by
$\cos\theta\equiv\hat{\omega}\cdot\hat{\mathbf{n}}$. The top graph
shows a highly peaked distribution around $\theta=0$ and
$\theta=\pi$ in the inset. However, the kernel is much less
anisotropic in this case. The lower graph has a purely quadrupolar
distribution, which reflects in the fact that the kernel the same
orientational dependence. In both cases, the kernel has a
\emph{minimum} at $\gamma=\pi/2$. Consequently, we expect no
isotropic--nematic transition. In the next section, we prove that
this conclusion holds for any choice of parameters.

\section{Bifurcation theory}
The ODF that minimizes the free energy \eqref{eq:F} obeys the
Euler-Lagrange equation
\begin{equation}
\ln[4\pi f(\hat{\omega})] = \lambda - \int{\rm
d}\hat{\omega}^{\prime}f(\hat{\omega}^{\prime})K(\hat{\omega},\hat{\omega}^{\prime}),
\end{equation}
where we introduced the Lagrange multiplier $\lambda$ to ensure
the normalization of $f$ given by Eq.~\eqref{eq:ODF:norm}. The
Euler-Lagrange equation can be rewritten in a form that always
satisfies this normalization
\begin{equation}\label{eq:EL}
f(\hat{\omega}) = \frac{\exp\bigl[-\int{\rm
d}\hat{\omega}^{\prime}f(\hat{\omega}^{\prime})K(\hat{\omega},\hat{\omega}^{\prime})\bigr]}{\displaystyle\int{\rm
d}\hat{\omega}^{\prime} \exp\biggl[-\int{\rm
d}\hat{\omega}^{\prime\prime}f(\hat{\omega}^{\prime\prime})K(\hat{\omega}^{\prime},\hat{\omega}^{\prime\prime})\biggr]}.
\end{equation}
One easily checks that $f_{\rm iso}(\hat{\omega})=1/4\pi$ is a
solution of Eq.~\eqref{eq:EL}, describing the perfectly isotropic
phase. Due to its nonlinear character, one can expect additional
(anisotropic) solutions to this equation. Finding explicit
expressions for these solutions, however, is difficult, although
good insight can be obtained form a bifurcation analysis. The goal
of this analysis is to determine if|and for what parameters|an
instability can be found in the reference solution with respect to
a perturbation.

We choose the isotropic ODF \eqref{eq:f-iso} as a reference, and
expand around this solution by writing
$f(\hat{\omega})=1/4\pi+\delta f(\hat{\omega})$, with $\delta f$ a
small deviation. Following the same scheme as \citet{Kayser}
applied to Onsager's model of hard rods|which was extended by
\citet{Mulder}|we find the bifurcation equation
\begin{equation}\label{eq:bif}
\delta f(\hat{\omega}) = -\int{\rm
d}\hat{\omega}^{\prime}K(\hat{\omega},\hat{\omega}^{\prime})\delta
f(\hat{\omega}^{\prime}) \equiv -\mathcal{K}[\delta
f](\hat{\omega}).
\end{equation}
This is an eigenvalue equation, for which a non-trivial solution
exists if the integral operator $\mathcal{K}$ has eigenvalues
$-1$. The parameter value for which this occurs is called the
bifurcation point, where an anisotropic solution branches off from
the (isotropic) reference solution. The solution to the
bifurcation equation \eqref{eq:bif} can be given in terms of
eigenfunctions of $\mathcal{K}$. On the basis of
rotational-symmetry arguments, we find that these eigenfunctions
are the Legendre polynomials of the dot product of the orientation
$\hat{\omega}$ with respect to an arbitrary direction,
\begin{equation}
\int{\rm
d}\hat{\omega}^{\prime}K(\hat{\omega},\hat{\omega}^{\prime})P_{\ell}(\hat{\omega}^{\prime}\cdot\hat{\mathbf{z}})
= \lambda_{\ell}P_{\ell}(\hat{\omega}\cdot\hat{\mathbf{z}}),
\end{equation}
where the eigenvalues $\lambda_{\ell}$ follow from
Eq.~\eqref{eq:K},
\begin{equation}
\lambda_{\ell} = \left\{\begin{array}{ll}
\displaystyle\frac{l_{\rm
B}}{\bar{\kappa}^{5}a^{2}R^{2}}\frac{4\pi\sigma_{\ell}^{2}}
{\Lambda_{\ell}(\epsilon;\bar{\kappa}a,\bar{\kappa}R)\Xi_{\ell}(\epsilon;\bar{\kappa}a,\bar{\kappa}R)}
& \mbox{for $\ell$ even},\\
0 & \mbox{for $\ell$ odd}.
\end{array}\right.
\end{equation}
The bifurcation point is determined by $\lambda_{\ell}=-1$.
However, all coefficients $\lambda_{\ell}$ are positive. Therefore
the bifurcation equation \eqref{eq:bif} has no solution, and there
is no bifurcation point. The understanding of the origin of this
property of the coefficients lies in the fact that both
$\Xi_{\ell}(\epsilon;\bar{\kappa}a,\bar{\kappa}R)$ and
$\Lambda_{\ell}(\epsilon;\bar{\kappa}a,\bar{\kappa}R)$ approach
their minimum in the limit $R\rightarrow a$. Moreover, these
limits are non-negative, since
\begin{equation}
\lim_{R\rightarrow
a}\Xi_{\ell}(\epsilon;\bar{\kappa}a,\bar{\kappa}R) =
\frac{1}{\bar{\kappa}^{2}a^{2}},
\end{equation}
\begin{equation}
\lim_{R\rightarrow
a}\Lambda_{\ell}(\epsilon;\bar{\kappa}a,\bar{\kappa}R) =
\frac{\epsilon\ell}{\bar{\kappa}^{3}a^{3}}.
\end{equation}
This result shows that the breaking of orientational symmetry
cannot be captured in this simple version of Poisson-Boltzmann
cell theory. Additionally, it strongly suggests that there must be
positional order before there can be orientational ordering in
suspensions of Janus or other patchy particles. In other words,
the transition from an isotropic state to a fully ordered crystal
phase|if it exists|will be intermitted by a plastic crystal phase.
Our simple cell model does not take into account the positional
correlations of the plastic crystal phase, due to the mean-field
nature of the applied boundary conditions at the cell surface.

\section{Conclusion and outlook}
We developed a simple cell model in the context of
Poisson-Boltzmann theory for heterogeneously charged colloidal
spheres. The boundary conditions|on the colloid surface as well as
on the Wigner-Seitz cell surface|depend on the charge
heterogeneity and the orientational distribution of the colloidal
particles. Within a linear approximation to Poisson-Boltzmann
theory, these boundary conditions give rise to a free-energy
functional of the orientational distribution function
$f(\hat{\omega})$ that is very similar to the one used in
Onsager's second-order virial approximation in the description of
the isotropic--nematic transition of hard rods \citep{Onsager}.
The present description, however, does \emph{not} give rise to
orientational ordering. Since our model treats the position of the
colloids in a mean field description|and since we do expect some
degree of orientational ordering at sufficiently high particle
density|this result suggests that orientational ordering requires
the existence of positional ordering. The present theory predicts
no orientational ordering in fluids of these particles, i.e., no
liquid crystal phases. On the other hand, this effect can also be
attributed to an oversimplification of our description. In that
sense, the orientational ordering could be the result of particle
pair correlations which are not included in our model.

Therefore, one could consider to expand the class of
Poisson-Boltzmann cell models even further. Conceivably, a method
can be devised to include correlations of particle positions
through the electrostatic boundary conditions. More specifically,
one can choose a different approach to the way that the surface
potential $\Phi_{R}$ is determined. In the present models, this
potential is independent of the colloidal species to which the
cell belongs. Also, each colloidal species has an equivalent
weight|equal to its molar fraction|in the average of the potential
and electric field flux at the cell boundary. This property is due
to the mean field description which is used. Also inherent to this
description is the assumption that the surrounding of a particle
at the cell boundary is independent from the species it belongs
to|or equivalently, its orientation. However, if this restriction
is lifted, one may include the fact that the surroundings do
depend on this property through the pair distribution function.

The nature of these correlations can be related in a simple way to
systems of oppositely charged colloidal particles
\citep{Caballero,Leunissen}. The number of bonds between
oppositely charged particles in these systems depends on the
colloid density. Also, for the dense liquid phase|coexisting with
a dilute vapor phase provided the Debye screening length is large
enough|the pair distribution function shows that a colloidal
particle is surrounded by different layers of colloidal species
with alternating signs of charge \citep{Caballero}. The first
surrounding layer has an opposite charge with respect to the
particle in the origin; the following layer is like-charged, and
so on. These systems also display multiple crystal structures,
which have different coordination number. The same notion can be
subsequently applied to particles of different orientations to
include orientational pair correlations in the cell model. This
paves the way to a description of orientationally ordered phases.
Additionally, a jellium approximation can be applied in the same
way it is applied to monodisperse systems of homogeneously charged
colloidal spheres and rods \citep{Trizac:jellium,Pianegonda}. In
this description, there is no need for a certain cell shape and
volume. Moreover, the jellium model has a natural way to include
particle pair correlations \citep{Castaneda,Colla}. Finally, there
is an opportunity to apply the Poisson-Boltzmann cell model to
non-spherical cells \citep{Graf}. The boundary conditions can be
imposed in the same way as in this paper. However, this
complicates the expression of the appropriate boundary conditions,
since a non-spherical shape will couple different spherical
harmonic modes. The shape of these cells must be controlled by
additional constraints, such as the minimization of free energy.
Also, the choice of shapes must be motivated by physical
arguments. We leave these options for future studies.

\appendix

\section{Derivation of the electrostatic potential}\label{sec:deriv}
Inside the colloidal particle $\Phi(\hat{\omega};\mathbf{r})$
satisfies the Laplace equation, whereas in the cell interior it
satisfies the LPB-equation. Therefore, we have to match two
general solutions, using the boundary conditions on the particle
surface. We do this by expanding both solutions in spherical
harmonics. These expressions are given in Eqs.~\eqref{eq:sol-in}
and \eqref{eq:sol-out}. We apply the boundary condition on the
particle surface, given by Eqs.~\eqref{eq:boun:col-V} and
\eqref{eq:boun:col-E}. To this end, we expand the surface charge
distribution in spherical harmonics, using
Eq.~\eqref{eq:charge-dist} and the addition theorem,
\begin{equation}
\sigma(\hat{\omega};\hat{\mathbf{n}}) =
\sum_{\ell=0}^{\infty}\sum_{m=-\ell}^{+\ell}\sigma_{\ell}Y_{\ell,m}^{*}(\hat{\omega})Y_{\ell,m}(\hat{\mathbf{n}}).
\end{equation}
The arguments $\hat{\mathbf{n}}$ and $\hat{\omega}$ of the
spherical harmonic function should be interpreted as the pair of
spherical angles of this orientation with respect to the reference
frame. Consequently, from the boundary conditions
\eqref{eq:boun:col-V} and \eqref{eq:boun:col-E}, we obtain the
following condition on the coefficients of $\Phi_{\rm out}$,
\begin{align}\label{eq:bouns}
& B_{\ell,m}(\hat{\omega})\left(i_{\ell}^{\prime}(\bar{\kappa}a) -
\frac{\epsilon\ell}{\bar{\kappa}a}i_{\ell}(\bar{\kappa}a)\right)\nonumber\\
{}+{}&
C_{\ell,m}(\hat{\omega})\left(k_{\ell}^{\prime}(\bar{\kappa}a) -
\frac{\epsilon\ell}{\bar{\kappa}a}k_{\ell}(\bar{\kappa}a)\right)\nonumber\\
&{}= -4\pi l_{\rm
B}\bar{\kappa}^{-1}\sigma_{\ell}Y_{\ell,m}^{*}(\hat{\omega}).
\end{align}

Next, we apply the boundary conditions at the cell surface given
in Eqs.~\eqref{eq:boun:cell-V-ani} and \eqref{eq:boun:cell-E-ani}.
This yields a linear system of equations, which can be solved
analytically. However, we can choose to split the solution into
two contributions. The first contribution then satisfies the
boundary conditions on the particle surface|given by
Eq.~\eqref{eq:bouns}|as well as the condition that the potential
vanishes at the cell boundary. This is already the relevant
boundary condition for all odd contributions to
$\Phi(\hat{\omega};\mathbf{r})$, whereas a second contribution
must be added later to the even contributions in order to satisfy
the full set of boundary conditions. The coefficients that belong
to the first contribution will be denoted by
$B_{\ell,m}(\hat{\omega})$ and $C_{\ell,m}(\hat{\omega})$. First,
we impose the vanishing potential at the cell boundary by
\begin{equation}
B_{\ell,m}(\hat{\omega})i_{\ell}(\bar{\kappa}R) +
C_{\ell,m}(\hat{\omega})k_{\ell}(\bar{\kappa}R) = 0.
\end{equation}
Together with Eq.~\eqref{eq:bouns}, this yields
\begin{align}\label{eq:B}
B_{\ell,m}(\hat{\omega}) ={}& -\frac{4\pi l_{\rm
B}\bar{\kappa}^{-1}\sigma_{\ell}Y_{\ell,m}^{*}(\hat{\omega})}{\Xi_{\ell}(\epsilon;\bar{\kappa}a,\bar{\kappa}R)}k_{\ell}(\bar{\kappa}R),\\\label{eq:C}
C_{\ell,m}(\hat{\omega}) ={}& \frac{4\pi l_{\rm
B}\bar{\kappa}^{-1}\sigma_{\ell}Y_{\ell,m}^{*}(\hat{\omega})}{\Xi_{\ell}(\epsilon;\bar{\kappa}a,\bar{\kappa}R)}i_{\ell}(\bar{\kappa}R),
\end{align}
where $\Xi_{\ell}(\epsilon;\bar{\kappa}a,\bar{\kappa}R)$ is
defined in Eq.~\eqref{eq:Xi}. The orientational dependence of this
first contribution is such that it|and therefore all odd
contributions|only depends on the angle between $\hat{\omega}$ and
$\hat{\mathbf{r}}$,
\begin{align}
\Phi_{\rm odd}(\hat{\omega};\mathbf{r}) ={}& l_{\rm
B}\bar{\kappa}^{-1}
\underbrace{\sum_{\ell=1}^{\infty}}_{\ell\;{\rm odd}}(2\ell+1)\sigma_{\ell}P_{\ell}(\hat{\omega}\cdot\hat{\mathbf{r}})\nonumber\\
&{}\times
\frac{k_{\ell}(\bar{\kappa}r)i_{\ell}(\bar{\kappa}R)-i_{\ell}(\bar{\kappa}r)k_{\ell}(\bar{\kappa}R)}
{\Xi_{\ell}(\epsilon;\bar{\kappa}a,\bar{\kappa}R)}.
\end{align}

As previously mentioned, a second contribution must be added to
the coefficients of the even contributions. With this contribution
included, the solution $\Phi(\hat{\omega};\mathbf{r})$ satisfies
the full set of boundary conditions on the cell surface given in
Eqs.~\eqref{eq:boun:cell-V-ani} and \eqref{eq:boun:cell-E-ani}. We
denote the coefficients of this secondary contribution by
$\tilde{B}_{\ell,m}$ and $\tilde{C}_{\ell,m}$. Also, we show that
these do not dependent on the particle orientation, because the
two distinct boundary conditions that govern them do not. First,
the boundary condition \eqref{eq:bouns} is already satisfied by
the coefficients $B_{\ell,m}(\hat{\omega})$ and
$C_{\ell,m}(\hat{\omega})$. Therefore,
\begin{align}\label{eq:tilde-col}
& \tilde{B}_{\ell,m}\left(i_{\ell}^{\prime}(\bar{\kappa}a) -
\frac{\epsilon\ell}{\bar{\kappa}a}i_{\ell}(\bar{\kappa}a)\right)\nonumber\\
{}+{}& \tilde{C}_{\ell,m}\left(k_{\ell}^{\prime}(\bar{\kappa}a) -
\frac{\epsilon\ell}{\bar{\kappa}a}k_{\ell}(\bar{\kappa}a)\right) =
0.
\end{align}
Second, the boundary condition \eqref{eq:boun:cell-V-ani} imposes
a value on the coefficients that only depends on the value of the
potential at the cell surface, which is necessarily independent of
the orientation $\hat{\omega}$. Hence,
\begin{equation}
\tilde{B}_{\ell,m}i_{\ell}(\bar{\kappa}R) +
\tilde{C}_{\ell,m}k_{\ell}(\bar{\kappa}R) = \phi_{\ell,m},
\end{equation}
where $\phi_{\ell,m}$ is defined by
\begin{equation}
\Phi_{R}(\hat{\mathbf{n}}) = \Phi_{0} - \tanh\Phi_{0} +
\underbrace{\sum_{\ell=0}^{\infty}}_{\ell\;{\rm
even}}\sum_{m=-\ell}^{+\ell}\phi_{\ell,m}Y_{\ell,m}(\hat{\mathbf{n}}).
\end{equation}
Together, these conditions yield
\begin{align}
\tilde{B}_{\ell,m} ={}& -\left(k_{\ell}^{\prime}(\bar{\kappa}a) -
\frac{\epsilon\ell}{\bar{\kappa}a}k_{\ell}(\bar{\kappa}a)\right)\frac{\phi_{\ell,m}}{\Xi_{\ell}(\epsilon;\bar{\kappa}a,\bar{\kappa}R)},\\
\tilde{C}_{\ell,m} ={}& \left(i_{\ell}^{\prime}(\bar{\kappa}a) -
\frac{\epsilon\ell}{\bar{\kappa}a}i_{\ell}(\bar{\kappa}a)\right)\frac{\phi_{\ell,m}}{\Xi_{\ell}(\epsilon;\bar{\kappa}a,\bar{\kappa}R)}.
\end{align}
Finally, the boundary condition \eqref{eq:boun:cell-E-ani} imposes
a vanishing value of the weighted average of the even
contributions to the electric field flux at the cell boundary.
This condition can be expressed in terms of a relation between the
coefficients $B_{\ell,m}$, $C_{\ell,m}$, $\tilde{B}_{\ell,m}$, and
$\tilde{C}_{\ell,m}$. By substituting the values given in
Eqs.~\eqref{eq:B} and \eqref{eq:C}, we arrive at
\begin{align}\label{eq:tilde-cell}
&\tilde{B}_{\ell,m}i_{\ell}^{\prime}(\bar{\kappa}R) +
\tilde{C}_{\ell,m}k_{\ell}^{\prime}(\bar{\kappa}R)\nonumber\\
={}& -\int{\rm d}\hat{\omega}f(\hat{\omega})
\bigl[B_{\ell,m}(\hat{\omega})i_{\ell}^{\prime}(\bar{\kappa}R) +
C_{\ell,m}(\hat{\omega})k_{\ell}^{\prime}(\bar{\kappa}R)\bigr]\nonumber\\
={}& \frac{4\pi l_{\rm
B}\bar{\kappa}^{-1}\sigma_{\ell}}{\Xi_{\ell}(\epsilon;\bar{\kappa}a,\bar{\kappa}R)\bar{\kappa}^{2}R^{2}}
\int{\rm d}\hat{\omega}f(\hat{\omega})Y_{\ell,m}^{*}(\hat{\omega})
\qquad\mbox{for $\ell$ even},
\end{align}
where we used
\begin{equation}
k_{\ell}(\bar{\kappa}R)i_{\ell}^{\prime}(\bar{\kappa}R) -
i_{\ell}(\bar{\kappa}R)k_{\ell}^{\prime}(\bar{\kappa}R) =
\frac{1}{\bar{\kappa}^{2}R^{2}} \qquad\forall\ell,
\end{equation}
which can be derived from standard identities for the modified
spherical Bessel functions. The conditions in
Eqs.~\eqref{eq:tilde-col} and \eqref{eq:tilde-cell} are sufficient
to derive similar expressions for $\tilde{B}_{\ell,m}$ and
$\tilde{C}_{\ell,m}$. However, the construction we use to derive
the solution to the even contributions enables us to show that the
cell surface potential $\Phi_{R}(\hat{\mathbf{n}})$ has the same
symmetry properties as the ODF (in addition to the fact that it is
composed purely of even contributions),
\begin{equation}
\phi_{\ell,m} = \frac{4\pi l_{\rm
B}\bar{\kappa}^{-1}\sigma_{\ell}}{\Lambda_{\ell}(\epsilon;\bar{\kappa}a,\bar{\kappa}R)\bar{\kappa}^{2}R^{2}}
\int{\rm
d}\hat{\omega}f(\hat{\omega})Y_{\ell,m}^{*}(\hat{\omega}),
\end{equation}
where $\Lambda_{\ell}(\epsilon;\bar{\kappa}a,\bar{\kappa}R)$ is
defined in Eq.~\eqref{eq:Lambda}. With this we readily obtain the
even contributions
\begin{align}
&\Phi_{\rm even}(\hat{\omega};\mathbf{r}) = l_{\rm
B}\bar{\kappa}^{-1}
\underbrace{\sum_{\ell=0}^{\infty}}_{\ell\;{\rm even}}\frac{(2\ell+1)\sigma_{\ell}}{\Xi_{\ell}(\epsilon;\bar{\kappa}a,\bar{\kappa}R)}\nonumber\\
&{}\times\biggl\{
\bigl[k_{\ell}(\bar{\kappa}r)i_{\ell}(\bar{\kappa}R)-i_{\ell}(\bar{\kappa}r)k_{\ell}(\bar{\kappa}R)\bigr]
P_{\ell}(\hat{\omega}\cdot\hat{\mathbf{r}})\nonumber\\
&\qquad{}+ \frac{\Xi_{\ell}(\epsilon;\bar{\kappa}a,\bar{\kappa}r)}
{\Lambda_{\ell}(\epsilon;\bar{\kappa}a,\bar{\kappa}R)\bar{\kappa}^{2}R^{2}}
\int{\rm
d}\hat{\omega}^{\prime}f(\hat{\omega}^{\prime})P_{\ell}(\hat{\omega}^{\prime}\cdot\hat{\mathbf{r}})\biggr\},
\end{align}
and we obtain the general solution for the dimensionless
electrostatic potential in the cell interior given in
Eq.~\eqref{eq:sol-V}.

\end{document}